# Higher-order Klein bottle topological insulator in three-dimensional acoustic crystals

Yu-Liang Tao[1*], Mou Yan[2,3*], Mian Peng[4], Qiang Wei[4], Zhenxing Cui[4], Shengyuan A. Yang[5], Gang Chen[2,3,4†], Yong Xu[1†]

[1]Center for Quantum Information, IIS, Tsinghua University, Beijing 100084, China
[2]Key Laboratory of Materials Physics of Ministry of Education, School of Physics and Laboratory of Zhongyuan Light, Zhengzhou University, Zhengzhou 450001, China
[3]Institute of Quantum Materials and Physics, Henan Academy of Sciences, Zhengzhou 450046, China
[4]State Key Laboratory of Quantum Optics and Quantum Optics Devices, Institute of Laser spectroscopy, Shanxi University, Taiyuan 030006, China
[5]Research Laboratory for Quantum Materials, Singapore University of Technology and Design, Singapore, 487372, Singapore

*These authors contributed equally to this work.
†Corresponding author. Email: chengang971@163.com; yongxuphy@tsinghua.edu.cn

Topological phases of matter are classified based on symmetries, with nonsymmorphic symmetries like glide reflections and screw rotations being of particular importance in the classification. In contrast to extensively studied glide reflections in real space, introducing space-dependent gauge transformations can lead to momentum-space glide reflection symmetries, which may even change the fundamental domain for topological classifications, e.g., from a torus to a Klein bottle. Here we discover a new class of three-dimensional (3D) higher-order topological insulators, protected by a pair of momentum-space glide reflections. It supports gapless hinge modes, as dictated by the quadrupole moment and Wannier Hamiltonians defined on a Klein bottle manifold, and we introduce two topological invariants to characterize this phase. Our predicted topological hinge modes are experimentally verified in a 3D-printed acoustic crystal, providing direct evidence for 3D higher-order Klein bottle topological insulators. Our results not only showcase the remarkable role of momentum-space glide reflections in topological classifications, but also pave the way for experimentally exploring physical effects arising from momentum-space nonsymmorphic symmetries.

# I. INTRODUCTION

Discovering novel topological phases of matter has been one of the major themes in condensed matter physics and material sciences [1-3]. These phases are categorized according to symmetries, such as internal symmetries or crystal symmetries [4]. Among crystal symmetries, nonsymmorphic symmetries play an important role in the classification [5-9]. Such symmetries involve fractional lattice translations in real space, whereas their actions in momentum space is always symmorphic. In the past decade, rapid technological advances in engineering tight-binding models with high controllability in metamaterials have aroused considerable interest in studying topological phases under gauge fields [10-22]. Especially, in time-reversal-invariant metamaterials, one can readily deploy $\mathbb{Z}_2$ gauge fields, i.e., hopping amplitudes endowed with phases of $\pm 1$. The gauge fluxes can modify the symmetry algebra projectively. Particularly, it was found that the nontrivial projective symmetry algebra between translation and (symmorphic) reflection in real space could give rise to a *momentum-space* glide reflection, which involves fractional translations in momentum space [20]. Such a symmetry can even change the fundamental domain from a Brillouin torus to a Brillouin Klein bottle (which is non-orientable).

Crystal symmetries also play a crucial role in the classification of higher-order topological phases (HOTPs), which support $(n-m)$-dimensional gapless edge modes with $1 < m \leq n$ for an *n*-dimensional system [23-37]. Since the symmorphic action of these symmetries in momentum space does not affect the orientability of the fundamental domain for topological classifications, previous HOTPs were constructed based on an orientable base space (e.g., a torus) [23-37]. Given the fundamental importance of base space orientability for topological classifications, it is crucial to investigate whether there exists a momentum-space glide reflection induced non-orientable base space in 3Ds, so that a completely new class of higher-order topological insulators (HOTIs) beyond the existing paradigm of topological classifications arises.

Here we report the theoretical prediction and experimental realization of a 3D higher-order Klein bottle topological insulator (HOKBTI) protected by a pair of momentum-space glide reflection symmetries. We find that these exotic symmetries impose a constraint on the quadrupole moment and the Wannier-sector polarizations, allowing us to introduce two $\mathbb{Z}_2$ invariants to characterize the system's topology from two different perspectives. Notably, we demonstrate that this 3D Klein bottle insulator supports gapless modes at one-dimensional (1D) hinges. These gapless modes are not chiral and thus are completely different from the chiral hinge modes in conventional 3D HOTI [29]. We further experimentally realize this novel topological state in acoustic crystals.

## II. THEORETICAL MODEL

We start by considering a generic tight-binding Hamiltonian in square lattices containing four degrees of freedom in each unit cell. With momentum-space glide reflection symmetries $m_x$ and $m_y$, the Bloch Hamiltonian in momentum space $\mathcal{H}(\boldsymbol{k})$ satisfies

$$m_x \mathcal{H}(\boldsymbol{k}) m_x^{-1} = \mathcal{H}(-k_x, k_y, k_z + \pi), \tag{1}$$

$$m_y \mathcal{H}(\boldsymbol{k}) m_y^{-1} = \mathcal{H}(k_x, -k_y, k_z + \pi), \tag{2}$$

where $\boldsymbol{k} = (k_x, k_y, k_z)$ is the Bloch wavevector. Let us consider a simple model shown in Fig. 1(a) to demonstrate how the symmetries appear. The model is constructed based on the 2D Benalcazar-Bernevig-Hughes (BBH) model [23], where there are four sites in each unit cell and each plaquette carries a $\pi$ flux. The 2D models are stacked into a 3D model by introducing hopping along $z$. To ensure that each vertical plaquette also carries a $\pi$ flux, we impose a phase of $-1$ on the hopping amplitude along $z$ for two diagonal sites in each unit cell (Fig. 1). Such a model respects the vertical reflections $\mathcal{M}_x$ and $\mathcal{M}_y$ satisfying $\{L_z, \mathcal{M}_\nu\} = 0$ ($\nu = x, y$) where $L_z$ is the unit translation along $z$. Specifically, for the particular gauge configuration (e.g., the one in Fig. 1), these reflection operations have to contain a complementary gauge transformation. As shown in Fig. 1(b) [Fig. 1(c)], after reflecting the lattice with respect to the $x$-normal ($y$-normal) plane, we need to further apply the gauge transformation $G_{M_x}(\boldsymbol{r}) = (-1)^z$ [$G_{M_y}(\boldsymbol{r}) = (-1)^{z+1} \tau_z \sigma_z$] [$\boldsymbol{r} = (x, y, z)$ is the position vector of sites] to transform it into the original lattice. The spatial dependence of the gauge transformation thus results in the anticommutative relation between reflection and translation. When considering their actions on momentum space, momentum-space glide reflections represented by $m_x = \tau_0 \sigma_1$ and $m_y = \tau_2 \sigma_2$ appear (see APPENDIX A for derivation). One can clearly see the effects of $m_x$ and $m_y$ if we explicitly write the Bloch Hamiltonian as

$$\mathcal{H}(\bm{k}) = (t_x + t'_x \cos k_x a_x)\Gamma_3 + t'_x \sin k_x a_x\, \Gamma_0 \\ +(t_y + t'_y \cos k_y a_y)\Gamma_1 + t'_y \sin k_y a_y\, \Gamma_2 + 2t_z \cos k_z a_z\, \Gamma_4, \quad (3)$$

where $\{\Gamma_j\}$ is a set of tensor products of Pauli matrices $\tau_\alpha$ and $\sigma_\beta$ (which act on internal degrees of freedom), specifically, $\Gamma_0 = \tau_0 \sigma_2$, $\Gamma_4 = \tau_0 \sigma_3$ and $\Gamma_i = \tau_i \sigma_1$ with $i = 1, 2, 3$, $t_\nu$ and $t'_\nu$ ($\nu = x, y$) are the intra-cell and inter-cell hopping strength along $\nu$, respectively, and $t_z$ is the hopping strength along $z$ [Fig. 1(a)] (we set $t_z = 0.5$ hereafter), and $a_\nu$ ($\nu = x, y, z$) is the lattice constant along $\nu$.

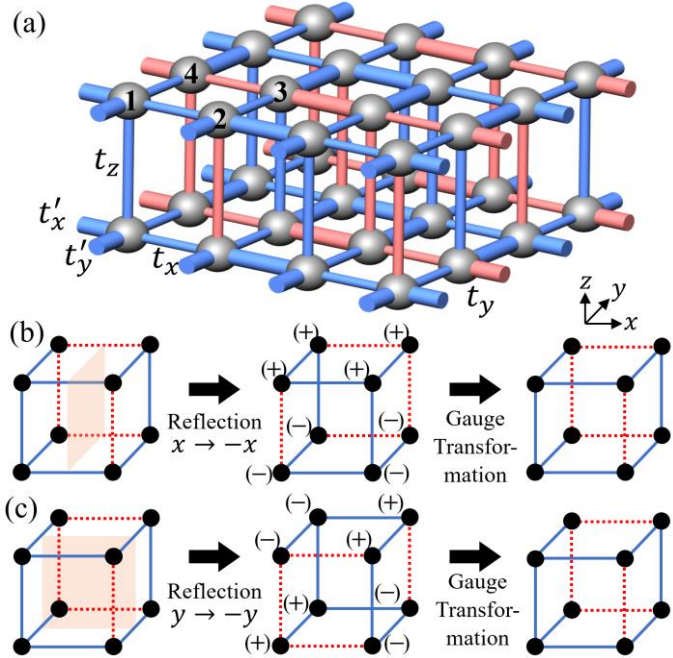

FIG. 1. (a) Schematic illustration of a tight-binding model for the HOKBTI. The red bonds indicate the hopping with the phase of -1 compared to the hopping through blue bonds. (b)-(c) Schematics of reflection symmetries with respect to $x$-normal (b) and $y$-normal (c) planes. In (b), the lattice is reflected about the red plane by $M_x$ followed

by applying a gauge transformation [$G_{M_x} = (-1)^z$]. The gauge transformation is described by imposing the phases in the brackets on the corresponding hopping. It follows that the combined operation $\mathcal{M}_x = G_{M_x} M_x$ does not change the hopping configuration, and thus the Hamiltonian remains invariant through this operation. In (c), $G_{M_y} = (-1)^{z+1} \tau_3 \sigma_3$.

For a generic Hamiltonian $\mathcal{H}(\boldsymbol{k})$, at a fixed $k_y$, the symmetry constraint in Eq. (1) changes the base space of a torus to a Klein bottle [similarly for the constraint in Eq. (2)] as shown in Fig. 2(a), similar to the 2D Klein bottle insulator [20]. We now analyze how the gapless hinge modes are topologically protected for the generic Hamiltonian from two different perspectives. For the first one, we consider the quadrupole moment, which is widely used to characterize the quadrupole insulator [38-41]. In APPENDIX B, we prove that the two symmetries impose a constraint on the quadrupole moment that

$$q_s = 0 \bmod 1, \tag{4}$$

where $q_s = q_{xy}(k_z) + q_{xy}(k_z + \pi)$ with $q_{xy}(k_z)$ being the quadrupole moment at the momentum $k_z$ along $z$. The relation tells us that $q_s$ can only take integer values. If we further require that $q_{xy}(k_z')$ is continuous as $k_z'$ changes from $k_z$ to $k_z + \pi$, then $q_s$ can take either 0 or 1 up to an integer multiple of 2 due to the gauge uncertainty. This enables us to define a $\mathbb{Z}_2$ index as

$$\chi_q = q_s \bmod 2, \tag{5}$$

which is quantized to 0 or 1. Note that $\chi_q$ is independent of $k_z$ since $\chi_q(k_z + \delta k_z)$ cannot suddenly change by one compared to $\chi_q(k_z)$ and thus has to be equal to $\chi_q(k_z)$.

The invariant identifies whether the quadrupole moment $q_{xy}(k_z')$ travels across 0.5 for an odd number of times as we change $k_z'$ from $k_z$ to $k_z + \pi$. Specifically, when $\chi_q = 1$, the value is crossed for an odd number of times [e.g., Fig. 2(c)]. While one can continuously deform the Hamiltonian to decrease the number of crossings, there always exists one crossing that cannot be removed due to the constraint in Eq. (4). As a consequence, one can always find a $k_{z0} \in [k_z, k_z + \pi]$ such that $q_{xy}(k_{z0}) = 0.5$, which guarantees the existence of gapless hinge modes. For example, for the specific simple model in Eq. (3), one finds two gapless branches in energy spectra describing the states spatially localized on a pair of diagonal hinges and the states localized on off-diagonal hinges, respectively [Fig. 2(d)]. The gapless hinge modes appear at $k_z = \pm \pi/2$, where $q_{xy} = 0.5$ [Fig. 2(c)]. In addition, due to the glide reflection symmetry, the spectrum in the range of $k_z \in [\pi, 2\pi)$ can be obtained by shifting the spectrum in the range of $k_z \in [0, \pi)$ by $\pi$. The hinge modes are clearly different from the chiral hinge modes in the conventional 3D HOTI [29], which are characterized by the winding number of the quadrupole moment [42,43]. Our topological state, as proved in APPENDIX B, does not possess any nonzero winding of the quadrupole moment.

When $\chi_q = 0$, the value of 0.5 is crossed for an even number of times [e.g., Fig. 2(c)]. In this case, we can continuously deform the Hamiltonian so that $q_{xy}(k_z) = 0$ for all $k_z$ and thus there do not exist gapless hinge modes protected by the quadrupole moment. Because of the constraint in Eq. (4), the topological invariant cannot change continuously and can only change abruptly when the energy gap closes, thereby serving as a well-defined topological invariant.

Secondly, we will show that the nontrivial higher-order Klein bottle topology can also be encoded in Wannier Hamiltonians, which have been widely used for studying higher-order topologies [23,24,29,37,44]. In the APPENDIX C, we prove that for a generic Hamiltonian respecting the two momentum-space glide reflection symmetries, the Wannier Hamiltonians $\mathcal{H}_{\mathcal{W}_y}$ and $\mathcal{H}_{\mathcal{W}_x}$ derived from occupied bands also preserve the two symmetries, that is, $m_x \mathcal{H}_{\mathcal{W}_y}(k_x, k_z) m_x^{-1} = \mathcal{H}_{\mathcal{W}_y}(-k_x, k_z + \pi)$ and $m_y \mathcal{H}_{\mathcal{W}_x}(k_y, k_z) m_y^{-1} = \mathcal{H}_{\mathcal{W}_x}(-k_y, k_z + \pi)$. The symmetry changes the base space from the Brillouin torus to the Brillouin Klein bottle, the grey region where the corresponding edges are glued together with the arrows matching [Fig. 2(b)]. Thus, $\mathcal{H}_{\mathcal{W}_y}$ and $\mathcal{H}_{\mathcal{W}_x}$ each is defined on a Klein bottle. In the following, we will establish the bulk-hinge correspondence based on the Wannier-sector polarization $p_\nu(k_z)$ ($\nu = x, y$) of a Wannier band, which represents the edge polarizations in a quadrupole insulator [23,24].

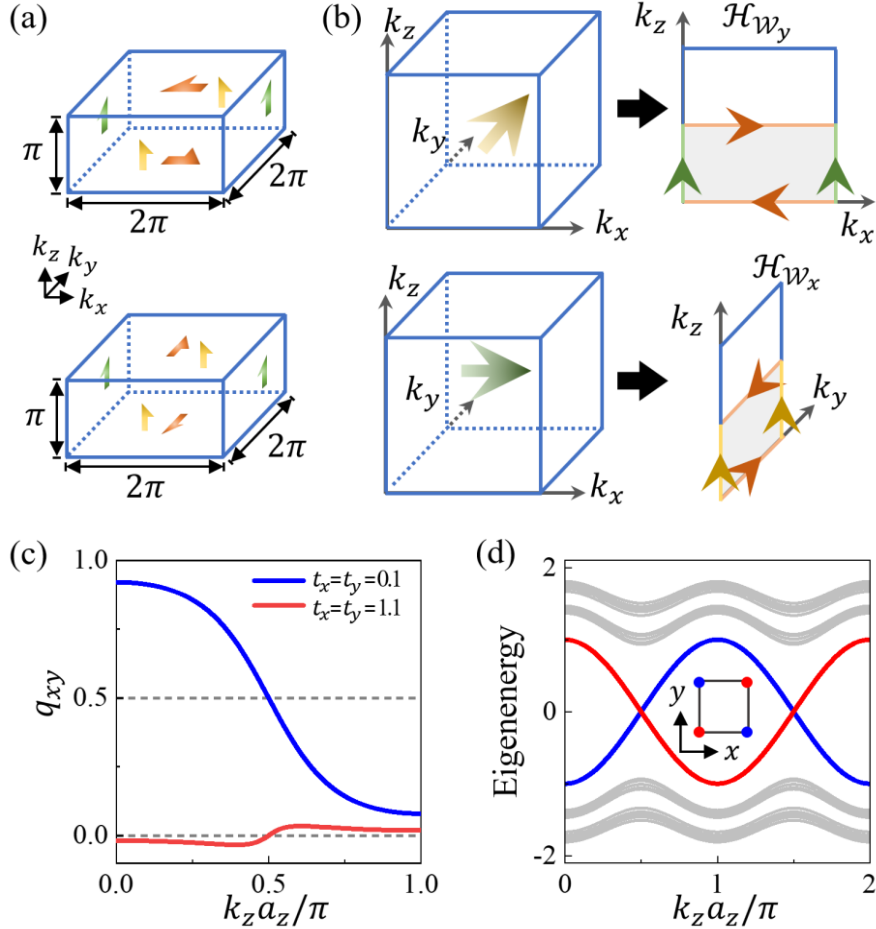

FIG. 2. (a) Momentum-space glide reflection symmetry, $m_x \mathcal{H}(\boldsymbol{k}) m_x^{-1} = \mathcal{H}(-k_x, k_y, k_z + \pi)$ (top) and $m_y \mathcal{H}(\boldsymbol{k}) m_y^{-1} = \mathcal{H}(k_x, -k_y, k_z + \pi)$ (bottom), resulting in the Brillouin Klein bottle at each slice with a fixed $k_y$ (top) or $k_x$ (bottom). (b) Schematic illustration of how the Klein bottle base space arises for the Wannier Hamiltonian. (c) The quadrupole moment $q_{xy}(k_z)$ versus $k_z$ for the Hamiltonian (3) with $t_x = t_y = 0.1$ (blue line) and $t_x = t_y = 1.1$ (red line). (d) Energy spectra versus $k_z$ for the Hamiltonian (3) at $t_x = t_y = 0.1$ with open and periodic boundaries in the $x$-$y$ plane and along $z$, respectively. The blue and red lines represent the hinge modes localized at diagonal hinges (solid blue circles) and off-diagonal hinges (solid red circles), respectively. In (c) and (d), $t'_x = t'_y = 1$.

In APPENDIX C, we prove that owing to the momentum-space glide reflections, such polarizations satisfy the relation $p_\nu(k_z) + p_\nu(k_z + \pi) = 0 \mod 1$ ($\nu = x, y$), ensuring that $\chi_\nu = [p_\nu(k_z) + p_\nu(k_z + \pi)] \mod 2$ is quantized to 0 or 1 [similar to $q_{xy}(k_z')$, $p_\nu(k_z')$ should be continuous from $k_z' = k_z$ to $k_z + \pi$]. When $\chi_\nu = 1$ ($\chi_\nu = 0$), the polarization $p_\nu(k_z')$ crosses 0.5 for an odd (even) number of times as $k_z'$ varies from $k_z$ to $k_z + \pi$. We then define a $\mathbb{Z}_2$-valued topological invariant as

$$\chi = \chi_x \chi_y. \tag{6}$$

When both $\chi_x$ and $\chi_y$ are equal to one, $\chi = 1$ so that the system is in a topologically nontrivial phase. For example, for the Hamiltonian in Eq. (3), when $|t_x| < 1$ and $|t_y| < 1$ (suppose that $t_x' = t_y' = 1$), both $p_x$ and $p_y$ cross 0.5 at $k_{z0} = \pm \pi/2$, leading to a topological phase with $\chi = 1$. In other regions of system parameters, the system is in a trivial phase with $\chi = 0$. For this specific simple model, in fact, $p_x(k_{z0}) = p_y(k_{z0}) = 0.5$, indicating that $\mathcal{H}(k_z = k_{z0})$ is effectively a quadrupole insulator with gapless corner modes [23]. In the 3D case, it implies the existence of gapless hinge modes. When more complicated terms are added to the model, one may not find a common $k_{z0}$ such that $p_x(k_{z0}) = p_y(k_{z0}) = 0.5$. However, the gapless hinge modes are still protected as long as $\chi = 1$ and the symmetries are respected (APPENDIX D).

The above analysis establishes the bulk-hinge correspondence for our 3D HOKBTI for a generic Hamiltonian from two different perspectives. For the concrete model in Eq. (3), the two topological invariants defined in Eq. (5) and Eq. (6) coincide, thereby giving the same phase diagram. We note that in some cases with long-range hoppings, anomalous phases arise which can only be characterized by the topological invariant $\chi_q$ (see APPENDIX D for discussions).

### III. RESULTS AND DISCUSSION

We now experimentally realize the state in an acoustic crystal. To achieve this, it is necessary to engineer the couplings between two acoustic cavities in such a way that each plaquette in the crystal carries a $\pi$ flux [45-47]. In our experiments, we fabricated identical air cuboid cavities that were coupled by narrow tubes using 3D-printing technology with photosensitive resin. The negative coupling in the $x$-$y$ plane was realized by shifting the connecting location [see Fig. 8(a)-(b) in APPENDIX E 1] compared to that for the positive coupling. Additionally, the positive and negative couplings along $z$ were achieved using tilted straight and bent tubes, respectively [see Fig. 8(c)-(d) in APPENDIX E 1]. Note that our experiment realized negative hoppings both in the $x$-$y$ plane and along $z$, in contrast to previous work where only the negative hopping in the $x$-$y$ plane was achieved [38-40]. Based on the above design, a 3D acoustic sample was manufactured to implement the tight-binding model in Eq. (3), with the cavities representing the lattice sites and the connecting tubes representing the hoppings. The fabricated acoustic crystal sample is shown in Fig. 3(a) which contains $4 \times 4 \times 21$ unit cells. In each unit cell, there are four cavities.

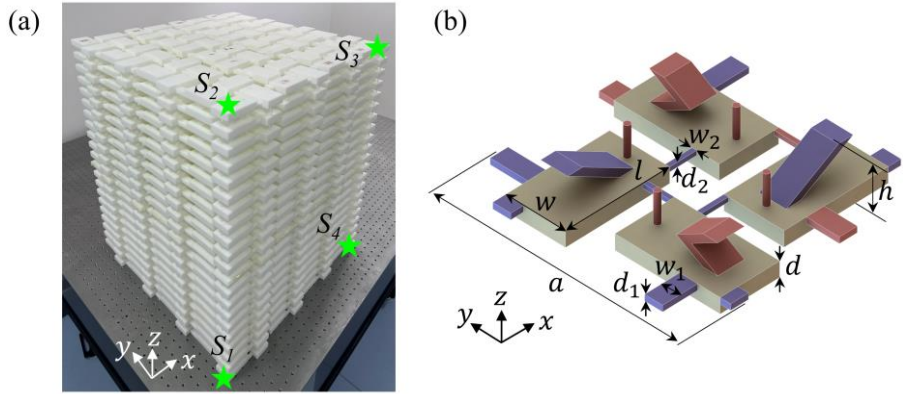

FIG. 3. (a) A photo of the 3D-printed acoustic crystal sample. The four green stars show the positions of four acoustic sources labelled by $S_1$, $S_2$, $S_3$ and $S_4$, respectively. (b) Schematic illustration of a unit cell for the acoustic crystal. The blue and red rectangular tubes between resonators represent the couplings with positive and negative signs, respectively.

To experimentally probe the hinge modes, we place an acoustic source at a bottom end of a hinge [position $S_1$ in Fig. 3(a)] and then use a detector to scan the acoustic signals in each cavity along the hinge. With the probed acoustic pressure field distribution, we perform the Fourier transform to obtain the acoustic dispersion with respect to $k_z$. The left panel of Fig. 4(a) shows the measured dispersion of the hinge states with positive group velocities around $k_z h = \pi/2$. The experimental results are consistent with the simulated results described by grey dots. Note that the simulated dispersions obtained by the commercial COMSOL Multiphysics solver package show the existence of gapless hinge modes, which agree well with our theoretical results in Fig. 2(d). To measure the dispersion with negative group velocities around $k_z h = 3\pi/2$ at the same hinge, we relocate the acoustic source to the top end of the same hinge (position $S_2$). The measured dispersion is displayed in the right panel of Fig. 4(a), which is in agreement with the simulated results. Furthermore, by placing the source at either the top or bottom end of a neighboring hinge (position $S_3$ and $S_4$), the measured results reveal the other branch of hinge modes localized at that specific hinge [Fig. 4(b)].

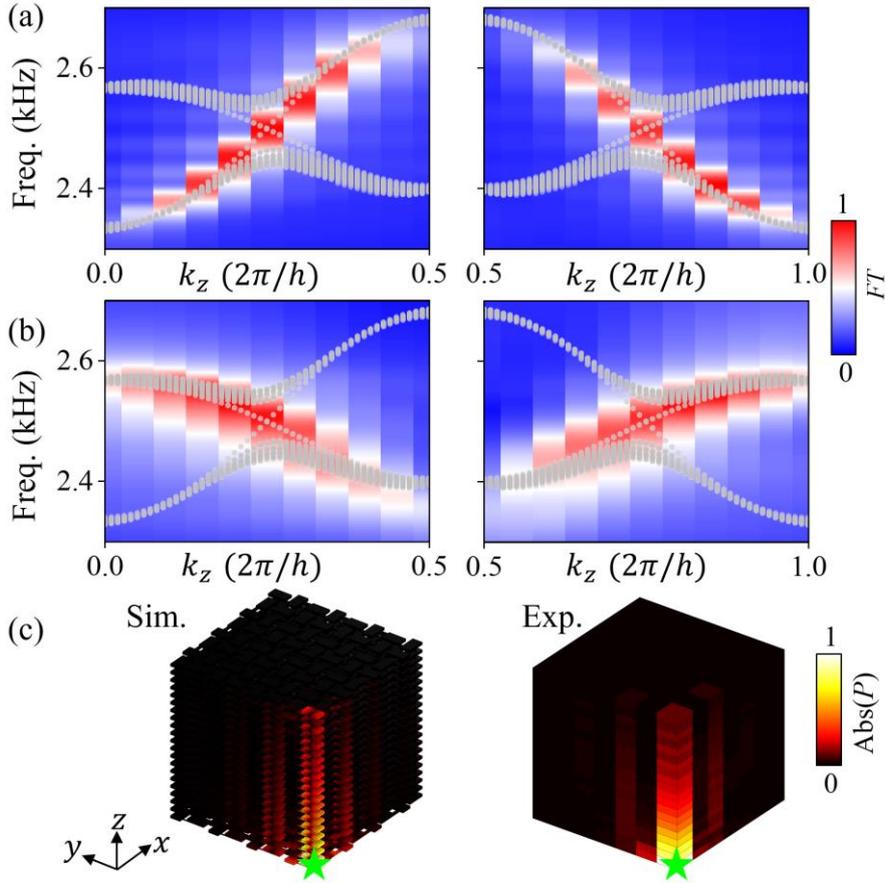

FIG. 4. (a), (b) Measured acoustic dispersions (colored) of the hinge states with positive or negative group velocities. The grey dots represent the simulated bulk and hinge

dispersions. (c) Simulated (left) and measured (right) acoustic pressure field distributions of the hinge states at the frequency of 2.5 kHz. In (c) and (d), the color bars are normalized to the maximums so that the value changes from 0 to 1.

In Fig. 4(c), we further present the simulated (left) and measured (right) acoustic pressure field distribution when an acoustic source with the frequency of 2.5 kHz is placed at the position $S_2$. In the simulation, we introduce a complex acoustic velocity given by $v = 343 * (1 + \alpha i)$ m/s with $\alpha = 0.007$ to account for system losses. Despite the presence of these losses, the propagation of an acoustic wave along the hinge is clearly observed, indicating the existence of hinge states. Our experimental results also confirm the propagation of signals along the hinge, which aligns well with the simulation results (see APPENDIX E 2 and 3 for further discussion on the effects of loss and the field distribution in the $x$-$y$ plane). Furthermore, in Fig. 12 of APPENDIX E 4, we show the experimental results at other frequencies outside the typical hinge mode range, where we observe that the acoustic wave can hardly propagate along the hinge. In Fig. 10 of APPENDIX E 2, we present the measured transmission spectrum, providing additional evidence for the presence of the hinge states.

## IV. CONCLUSIONS

In summary, we have theoretically proposed and experimentally observed a new type of HOTIs in a 3D acoustic crystal. Such a new topological state arises beyond the conventional classification because of the gauge field induced momentum-space nonsymmorphic symmetry. We introduce two topological invariants for this 3D HOKBTI. The phase is stable against weak disorder (see APPENDIX F). Our experiments further confirm the gapless hinge modes. Inspired by Refs. [48-49], we also find an anomalous HOKBTI where the gapless hinge modes only appear in entanglement spectra (see APPENDIX G). Our work thus provides an avenue for studying previously overlooked topological matters induced by gauge fluxes and exotic projective symmetry algebra.

## ACKNOWLEDGEMENTS

We thank J.-H. Wang for helpful discussions. This work was supported by the National Key R & D Program of China (Grant No. 2022YFA1404500), the National Natural Science Foundation of China (Grant Nos. 11974201, 12125406, 12074232, 12204290), Tsinghua University Dushi Program, the National Postdoctoral Program for Innovative Talents (Grant No. BX20220195), Cross-disciplinary Innovative Research Group Project of Henan Province (Grant No. 232300421004), and the Singapore MOE AcRF Tier 2 (T2EP50220-0026).

*Note added*. Recently, we became aware of two newly posted preprints [50,51] where first-order topological states arising from momentum-space nonsymmorphic symmetry are experimentally observed.

## APPENDIX A: GENERALIZED REFLECTION SYMMETRY WITH GAUGE FIELDS

In this section, we will present the generalized reflection symmetry realized by the composition of reflection and gauge transformations in both real and momentum space. In momentum space, the symmetry becomes a glide reflection symmetry.

### 1. Generalized reflection symmetry in real space

We write down a generic Hamiltonian in real space as

$$H = \sum_{r,d} \sum_{\alpha\beta} T_{\alpha\beta}(d) |r+d, \alpha\rangle\langle r, \beta|, \tag{S1}$$

where $T_{\alpha\beta}(d)$ denotes the hopping term from $|r, \beta\rangle$ to $|r+d, \alpha\rangle$ and $\alpha = 1, 2, 3, 4$ labels four degrees of freedom at each site. The two generalized reflection symmetries along $x$ and $y$ are defined by operators $\mathcal{M}_x$ and $\mathcal{M}_y$, respectively,

$$\mathcal{M}_x = G_{M_x} M_x, \tag{S2}$$

$$\mathcal{M}_y = G_{M_y} M_y, \tag{S3}$$

where $M_x$ and $M_y$ realize the traditional reflection of lattice sites along $x$ and $y$, respectively, that is, $M_x|\boldsymbol{r},\alpha\rangle = m'_x|-x,y,z,\alpha\rangle$ and $M_y|\boldsymbol{r},\alpha\rangle = m'_y|x,-y,z,\alpha\rangle$ with $\boldsymbol{r} = (x,y,z)$ labeling a unit cell and $m'_x$ and $m'_y$ reflecting internal degrees of freedom in a unit cell. For example, for the specific model in Eq. (3) in the main text, $m'_x$ and $m'_y$ reflect sites in a unit cell, that is, $m'_x|\boldsymbol{r},1\rangle = |\boldsymbol{r},2\rangle$, $m'_x|\boldsymbol{r},3\rangle = |\boldsymbol{r},4\rangle$, $m'_y|\boldsymbol{r},1\rangle = |\boldsymbol{r},4\rangle$ and $m'_y|\boldsymbol{r},2\rangle = |\boldsymbol{r},3\rangle$, where $\alpha$ labels a site in each unit cell (see Fig. 1 in the main text for our notation scheme). For the model, $G_{M_x} = (-1)^z$ and $G_{M_y} = (-1)^z G_0$ with $G_0 = -\tau_3\sigma_3$ realize the gauge transformations. Different from traditional gauge transformations, this gauge transformation is dependent on the position of unit cells. For a generic Hamiltonian, we summarize the operations as

$$\mathcal{M}_x|\boldsymbol{r},\alpha\rangle = (-1)^z m_x|-x,y,z,\alpha\rangle, \tag{S4}$$
$$\mathcal{M}_y|\boldsymbol{r},\alpha\rangle = (-1)^z m_y|x,-y,z,\alpha\rangle, \tag{S5}$$

where $m_x^2 = m_z^2 = 1$. For the specific model, $m_x = m'_x = \tau_0\sigma_1$ and $m_y = G_0 m'_y = \tau_2\sigma_2$.

Because of the position dependence of the gauge transformations, we can prove that

$$\{L_z, \mathcal{M}_\nu\} = 0, \tag{S6}$$

where $L_z$ is the translation operator along $z$ and $\nu = x, y$. The anti-commutation relation can be easily derived through the following calculations,

$$L_z\mathcal{M}_x|\boldsymbol{r},\alpha\rangle = (-1)^z L_z m_x|-x,y,z,\alpha\rangle = (-1)^z m_x|-x,y,z+1,\alpha\rangle, \tag{S7}$$
$$\mathcal{M}_x L_z|\boldsymbol{r},\alpha\rangle = -(-1)^z m_x|-x,y,z+1,\alpha\rangle = -L_z\mathcal{M}_x|\boldsymbol{r},\alpha\rangle. \tag{S8}$$

Similarly, one can obtain the anti-commutation relation between $L_z$ and $\mathcal{M}_y$.

We can also prove that Eq. (S3) and (S4) hold if the anti-commutation relation is satisfied. Specifically, we write the relation as

$$L_z\mathcal{M}_\nu L_z^{-1}\mathcal{M}_\nu^{-1} = -1, \tag{S9}$$

which leads to

$$-1 = L_z G_{M_\nu} M_\nu L_z^{-1} M_\nu^{-1} G_{M_\nu}^{-1}$$

$$= L_z G_{M_\nu} L_z^{-1} G_{M_\nu}^{-1}$$

$$= G_{M_\nu}(\boldsymbol{r} - \boldsymbol{d_z}) G_{M_\nu}^{-1}(\boldsymbol{r}). \tag{S10}$$

Here, $\nu = x, y$ and $\boldsymbol{d_z}$ is a unit vector along $z$. In the derivation, we have used the fact that $M_\nu L_z = L_z M_\nu$ and $L_z G_{M_\nu} L_z^{-1} = G_{M_\nu}(\boldsymbol{r} - \boldsymbol{d_z})$, which can be easily derived by applying the operator to $|\boldsymbol{r}, \alpha\rangle$. We thus derive that $G_{M_\nu}(\boldsymbol{r} - \boldsymbol{d_z}) = -G_{M_\nu}(\boldsymbol{r})$, which gives $G_{M_\nu}|\boldsymbol{r}, \alpha\rangle = (-1)^z (-1)^{\theta_{\nu\alpha}} |\boldsymbol{r}, \alpha\rangle$ with $\theta_{\nu\alpha}$ taking the values of 0 or 1. As a result, Eq. (S3) and (S4) are obtained.

## 2. Glide reflection symmetry in momentum space

In this subsection, we will present the formalism of $\mathcal{M}_x$ and $\mathcal{M}_y$ on the Hamiltonian in momentum space. We now write the tight-binding model Hamiltonian as

$$H = \sum_{\boldsymbol{k}} \sum_{\alpha\beta} [\mathcal{H}(\boldsymbol{k})]_{\alpha\beta} |\boldsymbol{k}, \alpha\rangle\langle\boldsymbol{k}, \beta|, \tag{S11}$$

where $|\boldsymbol{k}, \alpha\rangle$ with $\boldsymbol{k} = (k_x, k_y, k_z)$ denotes the momentum space basis, that is, $|\boldsymbol{k}, \alpha\rangle = \frac{1}{\sqrt{N}} \sum_{\boldsymbol{r}} e^{i\boldsymbol{k}\cdot\boldsymbol{r}} |\boldsymbol{r}, \alpha\rangle$ with $N$ being the total number of unit cells. Here, the matrix elements in $\mathcal{H}(\boldsymbol{k})$ is determined by

$$[\mathcal{H}(\boldsymbol{k})]_{\alpha\beta} = \sum_{\boldsymbol{d}} T_{\alpha\beta}(\boldsymbol{d}) e^{-i\boldsymbol{k}\cdot\boldsymbol{d}}. \tag{S12}$$

Now we consider the action of $\mathcal{M}_x$ on the hopping term with the site jump $\boldsymbol{d}$, which is expressed as

$$\mathcal{M}_x \sum_{\boldsymbol{r}} |\boldsymbol{r} + \boldsymbol{d}, \alpha\rangle\langle\boldsymbol{r}, \beta| \mathcal{M}_x^{-1}$$

$$= \sum_{\boldsymbol{r}} (-1)^{d_z} m_x |-x - d_x, y + d_y, z + d_z, \alpha\rangle\langle -x, y, z, \beta| m_x^{-1}$$

$$= \sum_{\boldsymbol{k}} e^{-i[-k_x d_x + k_y d_y + (k_z + \pi) d_z]} m_x |\boldsymbol{k}, \alpha\rangle\langle \boldsymbol{k}, \beta| m_x^{-1}. \tag{S13}$$

We see that the terms on $(k_x, k_y, k_z)$ are changed to the function of $(-k_x, k_y, k_z + \pi)$, leading to

$$\mathcal{M}_x H \mathcal{M}_x^{-1} = \sum_{\boldsymbol{k}} \sum_{\alpha\beta} [\mathcal{H}(-k_x, k_y, k_z + \pi)]_{\alpha\beta} m_x |\boldsymbol{k}, \alpha\rangle\langle \boldsymbol{k}, \beta| m_x^{-1}. \tag{S14}$$

Thanks to the $\mathcal{M}_x$ symmetry, the Hamiltonian satisfies

$$\mathcal{M}_x H \mathcal{M}_x^{-1} = H,$$

which gives rise to the glide reflection symmetry $m_x$ in momentum space, i.e.,

$$m_x \mathcal{H}(-k_x, k_y, k_z + \pi) m_x^{-1} = \mathcal{H}(\boldsymbol{k}). \tag{S15}$$

Similarly, one can derive the glide reflection symmetry $m_y$, that is,

$$m_y \mathcal{H}(k_x, -k_y, k_z + \pi) m_y^{-1} = \mathcal{H}(\boldsymbol{k}). \tag{S16}$$

### APPENDIX B: THE QUADRUPOLE MOMENT

In this section, we will prove that the quadrupole moment satisfies the relation $q_{xy}(k_z) + q_{xy}(k_z + \pi) = 0 \bmod 1$ enforced by two glide reflection symmetries, which ensures that the topological invariant is well defined. We also prove that the winding number of the quadrupole moment is zero. In a 2D $L_x \times L_y$ lattice system, the quadrupole moment $q_{xy}$ is defined as

$$q_{xy} = \frac{1}{2\pi} \text{Im}(\log \langle \hat{q}_{xy} \rangle), \tag{S17}$$

where

$$\hat{q}_{xy} = \exp\left[\frac{2\pi i}{L_x L_y} \sum_{r,\alpha} xy \left(\hat{c}^\dagger_{r,\alpha} \hat{c}_{r,\alpha} - \frac{1}{2}\right)\right], \tag{S18}$$

and $\hat{c}^\dagger_{r,\alpha}$ ($\hat{c}_{r,\alpha}$) is the fermion creation (annihilation) operator at the $\alpha$th site in the unit cell $\boldsymbol{r} = (x, y)$, and $\langle \hat{q}_{xy} \rangle$ is the expectation value of $\hat{q}_{xy}$ over the many-body ground state at half filing.

#### 1. The relation on the quadrupole moment

**Theorem .1.** *For a 3D Hamiltonian on a square lattice with translational symmetry, we define the quadrupole moment for a 2D system $H(k_z)$ at $k_z$ as*

$$q_{xy}(k_z) = \frac{1}{2\pi} \text{Im}(\log \langle \hat{q}_{xy} \rangle_{k_z}), \tag{S19}$$

*where $\langle \ldots \rangle_{k_z}$ calculates the expectation value over the many-body ground state of the 2D Hamiltonian $H(k_z)$ at a fixed $k_z$. If the 3D Hamiltonian respects $\mathcal{M}_x$ and $\mathcal{M}_y$ symmetry, then*

$$q_{xy}(k_z) + q_{xy}(k_z + \pi) = 0 \bmod 1. \tag{S20}$$

*Proof.* We first write down the Hamiltonian in the second quantization language,

$$\hat{H} = \sum_{k_z} \hat{H}(k_z) \tag{S21}$$

$$= \sum_{k_z} \sum_{r,r'} \sum_{\alpha,\alpha'} h_{r,\alpha;r',\alpha';k_z} \hat{c}^\dagger_{r,\alpha,k_z} \hat{c}_{r',\alpha',k_z}, \tag{S22}$$

where $\hat{c}^\dagger_{r,\alpha,k_z}$ ($\hat{c}_{r',\alpha',k_z}$) creates (annihilates) a fermionic particle (with momentum $k_z$ along $z$) of the $\alpha$th component in the site $r = (x,y)$ (here we view the sites in a unit cell as internal degrees of freedom), and $h$ is the hoping matrix.

When we apply generalized reflection operators to the creation and annihilation operators, we obtain

$$\widehat{\mathcal{M}}_x \hat{c}^\dagger_{r,\alpha,k_z} \widehat{\mathcal{M}}_x^{-1} = \hat{c}^\dagger_{D_{\widehat{\mathcal{M}}_x}(r,\alpha), k_z+\pi} \tag{S23}$$

$$\widehat{\mathcal{M}}_x \hat{c}_{r,\alpha,k_z} \widehat{\mathcal{M}}_x^{-1} = \hat{c}_{D_{\widehat{\mathcal{M}}_x}(r,\alpha), k_z+\pi} \tag{S24}$$

$$\widehat{\mathcal{M}}_y \hat{c}^\dagger_{r,\alpha,k_z} \widehat{\mathcal{M}}_y^{-1} = (-1)^{\theta_\alpha} \hat{c}^\dagger_{D_{\widehat{\mathcal{M}}_y}(r,\alpha), k_z+\pi} \tag{S25}$$

$$\widehat{\mathcal{M}}_y \hat{c}_{r,\alpha,k_z} \widehat{\mathcal{M}}_y^{-1} = (-1)^{\theta_\alpha} \hat{c}_{D_{\widehat{\mathcal{M}}_y}(r,\alpha), k_z+\pi} \tag{S26}$$

where $D_{\widehat{\mathcal{M}}_{x/y}}$ realize the real space transformation of $M_{x/y}$ on $(r,\alpha)$ in the 2D plane, and $\theta_\alpha$ takes the value of 0 or 1 (for the specific model in the main text, $\theta_1 = \theta_4 = 1$ and $\theta_2 = \theta_3 = 0$). The combination of $\widehat{\mathcal{M}}_x$ and $\widehat{\mathcal{M}}_y$ is a generalized rotational operator $\hat{\mathcal{R}} = \widehat{\mathcal{M}}_x \widehat{\mathcal{M}}_y$. Applying it to the creation and annihilation operators leads to

$$\hat{\mathcal{R}} \hat{c}^\dagger_{r,\alpha,k_z} \hat{\mathcal{R}}^{-1} = (-1)^{\theta_\alpha} \hat{c}^\dagger_{D_{\hat{\mathcal{R}}}(r,\alpha), k_z} \tag{S27}$$

$$\hat{\mathcal{R}} \hat{c}_{r,\alpha,k_z} \hat{\mathcal{R}}^{-1} = (-1)^{\theta_\alpha} \hat{c}_{D_{\hat{\mathcal{R}}}(r,\alpha), k_z} \tag{S28}$$

where $D_{\hat{\mathcal{R}}} = D_{\widehat{\mathcal{M}}_x} D_{\widehat{\mathcal{M}}_y}$.

Under the action of $\widehat{\mathcal{M}}_x$, $\hat{q}_{xy}(k_z)$ transforms as

$\widehat{\mathcal{M}}_x \hat{q}_{xy}(k_z) \widehat{\mathcal{M}}_x^{-1}$

$=\exp\left[\frac{2\pi i}{L_x L_y} \sum_{r,\alpha} xy \left(\hat{c}^\dagger_{D_{\widehat{\mathcal{M}}_x}(r,\alpha), k_z+\pi} \hat{c}_{D_{\widehat{\mathcal{M}}_x}(r,\alpha), k_z+\pi} - \frac{1}{2}\right)\right]$

$=\exp\left[\frac{2\pi i}{L_x L_y} \sum_{r,\alpha}(L_x + 2x_0 - 1 - x)y \hat{c}^\dagger_{r,\alpha,k_z+\pi} \hat{c}_{r,\alpha,k_z+\pi}\right] \exp\left(\frac{-2\pi i}{L_x L_y} \sum_r 2xy\right)$

$$
\begin{aligned}
=&\exp\left[-\frac{2\pi i}{L_xL_y}\sum_{r,\alpha}xy\left(\hat{c}^\dagger_{r,\alpha,k_z+\pi}\hat{c}_{r,\alpha,k_z+\pi}-\frac{1}{2}\right)\right]\\
&\times\exp\left[\frac{2\pi i(L_x+2x_0-1)}{L_xL_y}\sum_{r,\alpha}y\left(\hat{c}^\dagger_{r,\alpha,k_z+\pi}\hat{c}_{r,\alpha,k_z+\pi}-\frac{1}{2}\right)\right]\\
=&\,\hat{q}^\dagger_{xy}(k_z+\pi)\hat{p}_y(k_z+\pi).
\end{aligned}
\tag{S29}
$$

Here, $(x_0,y_0)$ is the position coordinate of the bottom left corner of the lattice, and $\hat{p}_y$ is a many-body order parameter about polarization, which is given by

$$
\hat{p}_y(k_z)=\exp\left[\frac{2\pi i(L_x+2x_0-1)}{L_xL_y}\sum_{r,\alpha}y\left(\hat{c}^\dagger_{r,\alpha,k_z}\hat{c}_{r,\alpha,k_z}-\frac{1}{2}\right)\right].
\tag{S30}
$$

Since $x_0$ is a fixed finite constant, $\hat{p}_y$ reduces to the standard definition of the polarization in the thermodynamic limit [52]. Under the action of $\hat{\mathcal{R}}$, $\hat{p}_y$ transforms as

$$
\begin{aligned}
\hat{\mathcal{R}}\hat{p}_y(k_z)\hat{\mathcal{R}}^{-1}&=\exp\left[\frac{2\pi i(L_x+2x_0-1)}{L_xL_y}\sum_{r,\alpha}y\left(\hat{c}^\dagger_{D_{\hat{\mathcal{R}}}(r,\alpha),k_z}\hat{c}_{D_{\hat{\mathcal{R}}}(r,\alpha),k_z}-\frac{1}{2}\right)\right]\\
&=\exp\left[-\frac{2\pi i(L_x+2x_0-1)}{L_xL_y}\sum_{r,\alpha}y\left(\hat{c}^\dagger_{r,\alpha,k_z}\hat{c}_{r,\alpha,k_z}-\frac{1}{2}\right)\right]\\
&=\hat{p}^\dagger_y(k_z),
\end{aligned}
\tag{S31}
$$

where we have considered the half-filled case.

Let $|\mathrm{GS}(k_z)\rangle$ be the ground state of $H(k_z)$ at half filling. Since the system respects the $\hat{\mathcal{M}}_x$ symmetry, $\hat{\mathcal{M}}_x|\mathrm{GS}(k_z)\rangle$ is the ground state of $H(k_z+\pi)$, that is, $|\mathrm{GS}(k_z+\pi)\rangle=\hat{\mathcal{M}}_x|\mathrm{GS}(k_z)\rangle$. The expectation value of $\hat{q}_{xy}(k_z+\pi)$ can be reduced as follows:

$$
\begin{aligned}
\langle\hat{q}_{xy}\rangle_{k_z+\pi}&=\langle\mathrm{GS}(k_z+\pi)|\hat{q}_{xy}(k_z+\pi)|\mathrm{GS}(k_z+\pi)\rangle\\
&=\langle\hat{\mathcal{M}}_x\mathrm{GS}(k_z)|\hat{q}_{xy}(k_z+\pi)|\hat{\mathcal{M}}_x\mathrm{GS}(k_z)\rangle\\
&=\langle\mathrm{GS}(k_z)|\hat{\mathcal{M}}^{-1}_x\hat{q}_{xy}(k_z+\pi)\hat{\mathcal{M}}_x|\mathrm{GS}(k_z)\rangle\\
&=\langle\mathrm{GS}(k_z)|\hat{q}^\dagger_{xy}(k_z)\hat{p}_y(k_z)|\mathrm{GS}(k_z)\rangle,
\end{aligned}
\tag{S32}
$$

where in the last step we have used Eq. S29. Although $|\mathrm{GS}(k_z)\rangle$ is generally not an eigenstate of $\hat{p}_y(k_z)$, we have the following result based on the perturbation theory [52,53]

$$
\hat{p}_y(k_z)|\mathrm{GS}(k_z)\rangle=\langle\hat{p}_y\rangle_{k_z}|\mathrm{GS}(k_z)\rangle+\mathcal{O}\left(\frac{1}{L_y}\right).
\tag{S33}
$$

Therefore, in the thermodynamic limit with an infinitely large $L_y$, we have

$$\langle \hat{q}_{xy} \rangle_{k_z+\pi} = \langle \hat{q}_{xy} \rangle^*_{k_z} \langle \hat{p}_y \rangle_{k_z}. \tag{S34}$$

Since the system also respects the $\hat{\mathcal{R}}$ symmetry, $\hat{\mathcal{R}}|\mathrm{GS}(k_z)\rangle$ is still the ground state of $H(k_z)$. Based on Eq. S31, the expectation value of $\hat{p}_y$ satisfies

$$\langle \hat{p}_y \rangle_{k_z} = \langle \mathrm{GS}(k_z)|\hat{p}_y(k_z)|\mathrm{GS}(k_z)\rangle$$

$$= \langle \hat{\mathcal{R}}\mathrm{GS}(k_z)|\hat{p}_y(k_z)|\hat{\mathcal{R}}\mathrm{GS}(k_z)\rangle$$

$$= \langle \mathrm{GS}(k_z)|\hat{\mathcal{R}}^{-1}\hat{p}_y(k_z)\hat{\mathcal{R}}|\mathrm{GS}(k_z)\rangle$$

$$= \langle \mathrm{GS}(k_z)|\hat{p}^\dagger_y(k_z)|\mathrm{GS}(k_z)\rangle$$

$$= \langle \hat{p}_y \rangle^*_{k_z}, \tag{S35}$$

which enforces $\langle \hat{p}_y \rangle_{k_z} = \pm 1$. To have a well-defined quadrupole moment, we require that the total polarization vanishes so that $\langle \hat{p}_y \rangle_{k_z} = 1$.

As a result, we arrive at

$$\langle \hat{q}_{xy} \rangle_{k_z+\pi} = \langle \hat{q}_{xy} \rangle^*_{k_z}. \tag{S36}$$

Based on the result, we obtain

$$q_{xy}(k_z) + q_{xy}(k_z + \pi) = \tfrac{1}{2\pi}\mathrm{Im}\big(\log\langle \hat{q}_{xy} \rangle_{k_z}\big) + \tfrac{1}{2\pi}\mathrm{Im}\big(\log\langle \hat{q}_{xy} \rangle_{k_z+\pi}\big)$$

$$= \tfrac{1}{2\pi}\mathrm{Im}\big(\log\langle \hat{q}_{xy} \rangle_{k_z}\big) + \tfrac{1}{2\pi}\mathrm{Im}\big(\log\langle \hat{q}_{xy} \rangle^*_{k_z}\big)$$

$$= 0 \bmod 1. \tag{S37}$$

### 2. The vanishing of the winding number of the quadrupole moment

We now prove that in the presence of a pair of momentum-space glide reflection symmetries, the winding number of the quadrupole moment vanishes.

Due to the constraint of the quadrupole moment in Eq. S20 enforced by momentum-space glide reflection symmetries, we have $q_{xy}(k_z) + q_{xy}(k_z + \pi) = n$ where $n$ an integer. The winding number of the quadruple moment is given by

$$W_q = \int_0^{2\pi} \frac{dq_{xy}(k_z)}{dk_z} dk_z = \int_0^{\pi} dq_{xy}(k_z) + \int_{\pi}^{2\pi} dq_{xy}(k_z) \tag{S38}$$

$$= \int_0^{\pi} dq_{xy}(k_z) + \int_0^{\pi} d[q_{xy}(k_z + \pi)] \tag{S39}$$

$$= \int_0^\pi dq_{xy}(k_z) + \int_0^\pi d[n - q_{xy}(k_z)] = 0. \tag{S40}$$

It indicates that the nontrivial gapless hinge modes in higher-order Klein bottle topological insulators are not chiral modes; the chiral modes are characterized by the nonzero $W_q$ [42,43].

### 3. Effects of terms breaking momentum-space glide reflection symmetries

In the proof in the subsection A, the momentum-space glide reflection symmetries are essential. Without these symmetries, $q_s = q_{xy}(k_z) + q_{xy}(k_z + \pi)$ is no longer quantized and the higher-order Klein bottle topology is not protected. To illustrate this, we add a term $H_\Delta = \Delta \tau_0 \sigma_3$ (which breaks both reflection symmetries) in the Hamiltonian (3) in the main text. Since $q_s$ is not required to be an integer, we can continuously vary $\Delta$ to remove the crossing of $q_{xy}(k_z)$ at 0.5 with respect to $k_z$ as shown in Fig. 5(a). As a result, the gapless hinge modes are removed through the adiabatic deformation [see the energy spectrum in Fig. 5(d)]. It is important to note that as we change $\Delta$, the bulk energy gap, the $x$-normal surface gap and $y$-normal surface gap remain open as shown in Fig. 5(b).

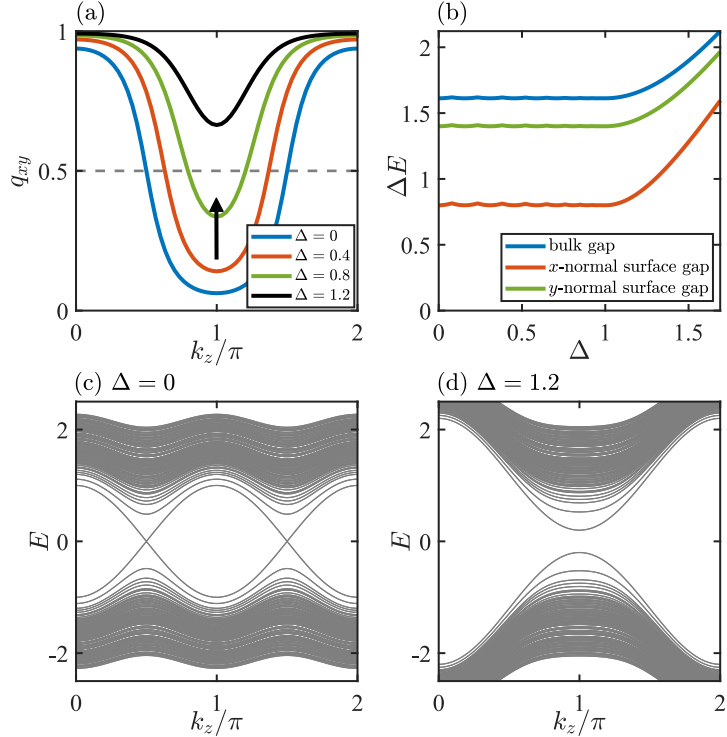

FIG. 5. Quadrupole moments, energy gaps and energy spectra for the Hamiltonian in Eq. (3) in the main text with the term $H_\Delta = \Delta\tau_0\sigma_3$. (a) The quadrupole moment $q_{xy}(k_z)$ versus $k_z$ for different $\Delta$. (b) The bulk energy gap (blue line), $x$-normal surface energy gap (red line) and $y$-normal surface energy gap (green line) versus $\Delta$. The energy spectrum with open boundaries along $x$ and $y$ and periodic boundary along $z$ for (c) $\Delta = 0$ and (d) $\Delta = 1.2$, respectively. We also set $a_x = a_y = a_z = 1$.

### APPENDIX C: WANNIER-SECTOR POLARIZATIONS

In this section, we will present two theorems regarding the Wannier Hamiltonian and Wannier-sector polarizations.

#### 1. Wannier band and Wannier Hamiltonian

First, we define the Wilson-loop operator $\mathcal{W}_y(k_x, k_{y0}, k_z)$ along the $k_y$-loop as [24]

$$\mathcal{W}_y(k_x, k_{y0}, k_z) = V^\dagger(k_x, k_{y0}, k_z)\left[\prod_{k_y}^{k_{y0}+2\pi \leftarrow k_{y0}} P(k_x, k_y, k_z)\right] V(k_x, k_{y0}, k_z), \quad \text{(S41)}$$

where $V(\boldsymbol{k}) = (|u_{\boldsymbol{k}}^1\rangle, |u_{\boldsymbol{k}}^2\rangle)$ is the matrix consisting of two occupied eigenvectors of $\mathcal{H}(\boldsymbol{k})$, and $P(\boldsymbol{k}) = V(\boldsymbol{k})V^\dagger(\boldsymbol{k})$ is the corresponding projection operator. Since $\mathcal{W}_y$ is a unitary matrix in the thermodynamic limit, its eigenvalue takes the form of $e^{iv_y^j}$ with $j = \pm$ labeling two bands. The corresponding eigenstate is $|v_y^j(\boldsymbol{k})\rangle$, that is,

$$\mathcal{W}_y(\boldsymbol{k})|v_y^j(\boldsymbol{k})\rangle = e^{iv_y^j}|v_y^j(\boldsymbol{k})\rangle \quad \text{(S42)}$$

with $\boldsymbol{k} = (k_x, k_{y0}, k_z)$. We thus define $v_y^\pm$ as a function of $(k_x, k_z)$ as the $y$-Wannier bands. Note that while $|v_y^\pm(\boldsymbol{k})\rangle$ may be dependent on the initial point $k_{y0}$, $v_y^\pm$ is independent of it [24].

Second, we define the $y$-Wannier band basis as

$$|w_y^\pm(\boldsymbol{k})\rangle = \sum_{n=1,2} |u_{\boldsymbol{k}}^n\rangle \left[v_{y,\boldsymbol{k}}^\pm\right]^n \quad \text{(S43)}$$

with $\left[v_{y,\boldsymbol{k}}^\pm\right]^n$ being the $n$th element in the eigenstate $|v_y^\pm(\boldsymbol{k})\rangle$. Thus, we define the $y$-Wannier Hamiltonian $\mathcal{H}_{\mathcal{W}_y}(k_x, k_z)$ based on Eq. 17 as

$$\mathcal{H}_{\mathcal{W}_y}(k_x, k_z) = \sum_{j=\pm} v_y^j(k_x, k_z) |w_y^j(\boldsymbol{k})\rangle\langle w_y^j(\boldsymbol{k})|. \tag{S44}$$

In the following, we will prove that the Wannier Hamiltonian obeys the glide reflection symmetry in momentum space.

**Theorem .2.** *If the system Hamiltonian $\mathcal{H}(\boldsymbol{k})$ respects two glide reflection symmetries $m_x$ and $m_y$, then the Wannier Hamiltonian $\mathcal{H}_{\mathcal{W}_y}$ and $\mathcal{H}_{\mathcal{W}_x}$ also respect the corresponding symmetry, that is,*

$$m_x \mathcal{H}_{\mathcal{W}_y}(k_x, k_z) m_x^{-1} = \mathcal{H}_{\mathcal{W}_y}(-k_x, k_z + \pi), \tag{S45}$$

$$m_y \mathcal{H}_{\mathcal{W}_x}(k_y, k_z) m_y^{-1} = \mathcal{H}_{\mathcal{W}_x}(-k_y, k_z + \pi). \tag{S46}$$

*Proof.* Since the Hamiltonian $\mathcal{H}(\boldsymbol{k})$ respects the glide reflection symmetry along $x$, $m_x V(\boldsymbol{k})$ is the set of occupied eigenvectors of $\mathcal{H}(-k_x, k_y, k_z + \pi)$. Inserting the identity $I = m_x^{-1} m_x$ into Eq. S41, we obtain

$$\mathcal{W}_y(k_x, k_z) \tag{S47}$$

$$= V^{\dagger}(k_x, k_{y0}, k_z) \left[ \prod_{k_y}^{k_{y0}+2\pi \leftarrow k_{y0}} m_x^{-1} m_x P(k_x, k_y, k_z) \right] m_x^{-1} m_x V(k_x, k_{y0}, k_z) \tag{S48}$$

$$= B_{m_x}^{\dagger} V^{\dagger}(\boldsymbol{k}') m_x^{-1} \left[ \prod_{k_y}^{k_{y0}+2\pi \leftarrow k_{y0}} P(-k_x, k_y, k_z + \pi) \right] V(\boldsymbol{k}') B_{m_x} \tag{S49}$$

$$= B_{m_x}^{\dagger} \mathcal{W}_y(-k_x, k_z + \pi) B_{m_x}, \tag{S50}$$

where $\boldsymbol{k}' = (-k_x, k_{y0}, k_z + \pi)$ and $B_{m_x} = V^{\dagger}(-k_x, k_{y0}, k_z + \pi) m_x V(k_x, k_{y0}, k_z)$ is a sewing matrix. This leads to $v_y^{\pm}(k_x, k_z) = v_y^{\pm}(-k_x, k_z + \pi)$ and $|v_y^{\pm}(-k_x, k_{y0}, k_z + \pi)\rangle = B_{m_x} |v_y^{\pm}(k_x, k_{y0}, k_z)\rangle$. The $y$-Wannier band basis thus satisfies

$$m_x |w_y^{\pm}(\boldsymbol{k})\rangle = m_x V(\boldsymbol{k}) |v_y^{\pm}(\boldsymbol{k})\rangle$$

$$= V(-k_x, k_{y0}, k_z + \pi) B_{m_x} |v_y^{\pm}(\boldsymbol{k})\rangle$$

$$= V(-k_x, k_{y0}, k_z + \pi) |v_y^{\pm}(-k_x, k_{y0}, k_z + \pi)\rangle$$

$$= |w_y^{\pm}(-k_x, k_{y0}, k_z + \pi)\rangle. \tag{S51}$$

As a result, we arrive at Eq. S45, implying that the topology of the $y$-Wannier Hamiltonian $\mathcal{H}_{\mathcal{W}_y}(k_x, k_z)$ is characterized over the topological domain of the Klein bottle, as shown in Fig. 2(b) in the main text. We note that while $\mathcal{H}_{\mathcal{W}_y}(k_x, k_z)$ also depends on $k_{y0}$, the topology is completely determined by $\mathcal{H}_{\mathcal{W}_y}$ at any fixed $k_{y0}$. This is attributable to the fact that the eigenvalues $v_y^\pm$ are independent of $k_{y0}$ and thus there are no topological phase transitions as we vary $k_{y0}$.

Similarly, based on the Wilson-loop operator $\mathcal{W}_x(k_{x0}, k_y, k_z)$ along the $k_x$-loop, we define the $x$-Wannier band basis $|w_x^\pm(\mathbf{k})\rangle$ and the $x$-Wannier Hamiltonian $\mathcal{H}_{\mathcal{W}_x}(k_y, k_z)$. Based on the glide reflection symmetry along $y$, we can also derive Eq. S46.

### 2. The relation on Wannier-sector polarizations

**Theorem .3.** *The Wannier-sector polarization of the $y$-Wannier band (similarly for the $x$-Wannier band) is defined as*

$$p_x(k_z) = -\frac{1}{(2\pi)^2}\int_{BZ} dk_x dk_y \mathcal{A}_{x,\mathbf{k}}^{v_y^-}, \qquad (S52)$$

*where $\mathcal{A}_{x,\mathbf{k}}^{v_y^-} = -i\langle w_y^-(\mathbf{k})|\partial_{k_x}|w_y^-(\mathbf{k})\rangle$ is the Berry connection over the $y$-Wannier bands $v_y^-$. If the system Hamiltonian $\mathcal{H}(\mathbf{k})$ respects two glide reflection symmetries $m_x$ and $m_y$, then*

$$p_v(k_z) + p_v(k_z + \pi) = 0 \text{ mod } 1, \qquad (S53)$$

*with $v = x, y$.*

*Proof.* Based on Eq. S51, we derive

$[p_x(k_z) + p_x(k_z + \pi)] \text{ mod } 1$

$= \left\{\frac{i}{(2\pi)^2}\int_{BZ} dk_x dk_y [\langle w_y^-(\mathbf{k})|\partial_{k_x}|w_y^-(\mathbf{k})\rangle + \langle w_y^-(\mathbf{k}_1)|\partial_{k_x}|w_y^-(\mathbf{k}_1)\rangle]\right\} \text{ mod } 1$

$= \left\{\frac{i}{(2\pi)^2}\int_{BZ} dk_x dk_y [\langle w_y^-(\mathbf{k})|\partial_{k_x}|w_y^-(\mathbf{k})\rangle + \langle w_y^-(\mathbf{k}_2)|m_x^{-1}\partial_{k_x}m_x|w_y^-(\mathbf{k}_2)\rangle]\right\} \text{ mod } 1$

$= \left\{\frac{i}{(2\pi)^2}\int_{BZ} dk_x dk_y [\langle w_y^-(\mathbf{k})|\partial_{k_x}|w_y^-(\mathbf{k})\rangle + \langle w_y^-(\mathbf{k}_2)|\partial_{k_x}|w_y^-(\mathbf{k}_2)\rangle]\right\} \text{ mod } 1$

$= 0, \qquad (S54)$

where $\mathbf{k}_1 = (k_x, k_y, k_z + \pi)$ and $\mathbf{k}_2 = (-k_x, k_y, k_z)$. Similarly, one can also derive that

$$[p_y(k_z) + p_y(k_z + \pi)] \mod 1 = 0. \tag{S55}$$

### APPENDIX D: EFFECTS OF LONG-RANGE HOPPINGS

In the main text, we consider the Hamiltonian with only nearest-neighbor intercell hopping terms that respect two glide reflection symmetries. There, the system reduces to the 2D BBH model at $k_z = \pm\pi/2$ where $p_x(k_z) = p_y(k_z) = 0.5$. In this section, we will show that this feature is not generic, and without it, the topological phase is still well protected. While long-range hopping terms are not present in our experiments, we will include some long-range hopping terms that respect the two glide reflection symmetries to illustrate the fact for theoretical interest. In addition, we will demonstrate the existence of anomalous phases which can only be characterized by $\chi_q$ by studying a model with long-range hoppings.

First, we add the following three next-nearest-neighbor intercell hopping terms in the Hamiltonian (3) in the main text,

$$\mathcal{H}_{\text{NNN}}(\boldsymbol{k}) = t_1 \sin k_z a_z \cos k_y a_y \, \tau_0 \sigma_3 + t_2 \sin k_z a_z \sin k_x a_x \, \tau_1 \sigma_0$$
$$+ t_3 \sin k_z a_z \sin k_y a_y \, \tau_1 \sigma_3, \tag{S56}$$

which respects the two glide reflection symmetries. With these terms, we cannot reduce the model to the BBH model at a certain $k_z$. However, we find that the presence of these terms does not change the phase diagram identified by the topological invariants $\chi$ and $\chi_q$. The topological phase also manifests in the presence of gapless hinges modes as shown in Fig. 6(a). The gapless modes for this set of system parameters arise at $k_z = 0.4808\pi$, which is slightly different from the case with $k_z = 0.5\pi$ for the model in the main text. In fact, at this $k_z$, $q_{xy} = 0.5$ [see Fig. 6(b)]. In other words, if $\chi_q = 1$ ($q_{xy}$ crosses 0.5 for odd number of times as we change $k_z$ from $-\pi$ to 0), then there must exist a $k_z$ where gapless hinge modes exist. In this case, although the topological invariant $\chi$ defined using Wannier-sector polarizations is equal to one, $p_x(k_z)$ and $p_y(k_z)$ do not cross 0.5 at the same $k_z$ [see Fig. 6(c)].

Second, we consider a model with some next-next-nearest-neighbor intercell hopping terms and demonstrate that involving these terms may render $\chi$ inaccurate

[44]. However, in that case, $\chi_q$ is still correct. The model reads

$$\mathcal{H}_2(\boldsymbol{k}) = (g_0 - \gamma - t_x - t_x\cos k_x + b_2 \cos 2k_y)\tau_3\sigma_2 - t_x \sin k_x \, \tau_3\sigma_3$$
$$+(\gamma + t_y \cos k_y)\tau_1\sigma_0 + (t_y \sin k_y + b_2 \sin 2k_y)\tau_3\sigma_1 + 2t_z \cos k_z \, \tau_2\sigma_0, \quad \text{(S57)}$$

where we have set $a_x = a_y = a_z = 1$. The Hamiltonian respects two momentum-space glide reflection symmetries $m_x = \tau_1\sigma_3$ and $m_y = \tau_1\sigma_1$, i.e., $m_x\mathcal{H}_2(\boldsymbol{k})m_x^{-1} = \mathcal{H}_2(-k_x, k_y, k_z + \pi)$ and $m_y\mathcal{H}_2(\boldsymbol{k})m_y^{-1} = \mathcal{H}_2(k_x, -k_y, k_z + \pi)$.

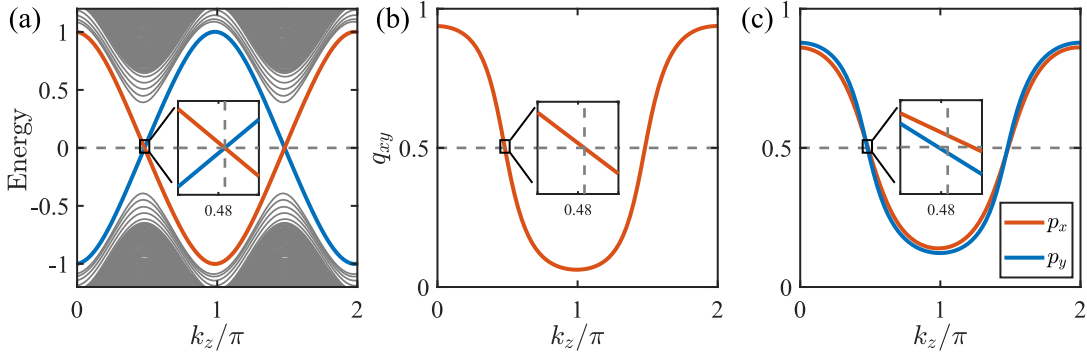

FIG. 6. Energy spectra, quadrupole moment and polarizations for the model with long-range hoppings. (a) Then energy spectrum with respect to $k_z$ for a system with open boundaries along $x$ and $y$ and periodic boundary along $z$. The blue line represents the hinge modes localized at the upper left and lower right hinges, and the red line represents the hinge modes localized at the upper right and lower left hinges [also see Fig. 2(d) in the main text]. (b) The quadrupole moment $q_{xy}$ as a function of $k_z$. (c) The Wannier-sector polarizations $p_x$ (red line) and $p_y$ (blue line) with respect to $k_z$. The insets are the zoomed-in view in the interval $k_z \in [0.475\pi, 0.485\pi]$. Here, $t_x = 0.3$, $t_y = 0.6$, $t'_x = t'_y = 1$, $t_z = 0.5$, $t_1 = 0.1$, $t_2 = 0.4$ and $t_3 = 0.4$. We also set $a_x = a_y = a_z = 1$.

Figure 7(a) displays $\chi$ and $\chi_q$ with respect to $\gamma$. We see that the equivalence between $\chi$ and $\chi_q$ is violated in the grey regions. In fact, it is the quadrupole moment $\chi_q$ that correctly describes the higher-order Klein bottle topological state. This is confirmed by Fig. 7(b) where the gapless hinge modes only exist in the region with $\chi_q = 1$. We further plot the energy spectrum with respect to $k_z$ at $\gamma = -0.75$ and $\gamma = 0.05$ in Figs. 7(c) and 7(d), respectively. We see the existence of gapless hinge modes in the latter with $\chi_q = 1$, irrespective of the value of $\chi$.

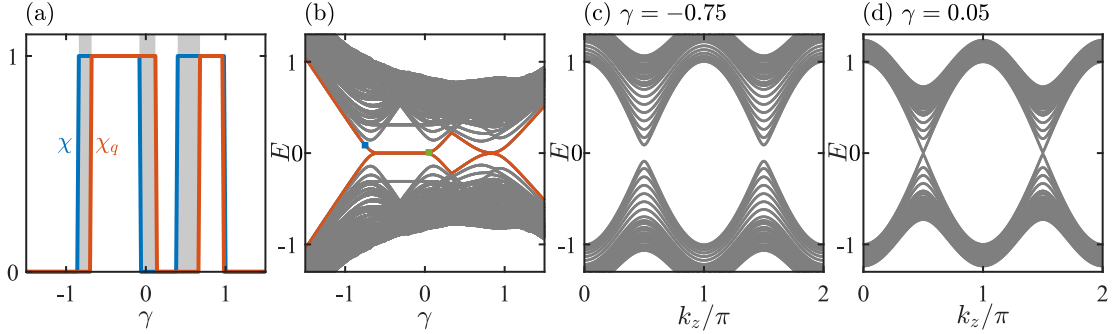

FIG. 7. (a) The phase diagram of $\chi$ and $\chi_q$ as a function of $\gamma$ for the Hamiltonian in Eq. (S57). In the grey regions, $\chi$ and $\chi_q$ are not equal. (b) The energy spectrum versus $\gamma$ for the system under open boundary conditions along $x$ and $y$ and periodic boundary conditions along $z$. The energy spectrum versus $k_z$ at (c) $\gamma = -0.75$ and (d) $\gamma = 0.05$. Here, $b_2 = 1.2$, $g_0 = 0.65$, $t_x = 1$, $t_y = 1$, and $t_z = 0.5$.

## APPENDIX E: EXPERIMENTAL DISCUSSIONS
### 1. Design of negative and positive couplings

In this section, we will provide the experimental parameters. Specifically, each unit cell shown in Fig. 3(b) in the main text contains four cavities with the length $l = 70$ mm, width $w = 40$ mm, and thickness $d = 10$ mm. Other parameters are $w_1 = 12$ mm, $d_1 = 4$ mm, $w_2 = 3$ mm and $d_2 = 3$ mm. The lattice constants in the $x$-$y$ plane and along $z$ are $a = 160$ mm and $h = 30$ mm, respectively. The tilted and bent tubes connecting neighboring layers along $z$ have the same cross section ($18 \times 10$ mm$^2$).

In addition, we will illustrate the design of negative or positive couplings between neighboring cavities. For a connected double cavity system, the Hamiltonian $H = \begin{pmatrix} f_0 & \kappa \\ \kappa & f_0 \end{pmatrix}$, where $f_0$ is the eigenfrequency of a single cavity and $\kappa$ is the coupling strength of the two cavities. Two split eigenmodes thus occur at frequencies of $f_\pm = f_0 \pm \kappa$ corresponding to the eigenvectors $\frac{1}{\sqrt{2}}(1, \pm 1)^T$; the eigenvectors with the sign

+ and − represent the in-phase and out-of-phase mode, respectively. When $\kappa > 0$ ($\kappa < 0$), the in-phase mode has higher (lower) frequency than the out-of-phase mode. In Fig. 8, we provide our numerical simulation results for different linking schemes. We see that the out-of-phase dipole mode has higher frequency than the in-phase one for the linking schemes in Figs. 8(b) and 8(d), indicating that the coupling between the two cavities is negative. In contrast, the out-of-phase mode has lower frequency than the in-phase one for the linking schemes in Figs. 8(c) and 8(e), implying that the coupling is positive.

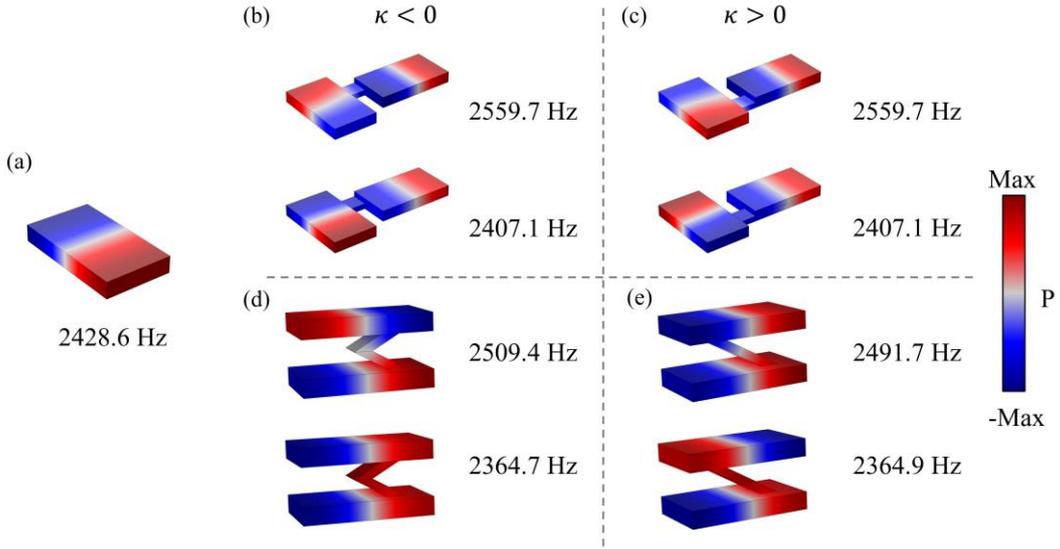

FIG. 8. Simulated results of the positive and negative couplings. (a) The dipole mode supported in a single cavity. (b)-(e), Acoustic pressure field distributions for different linking structures. One can also find the the resonant frequency and type of of the dipole mode beside the structure.

To demonstrate the positive or negative couplings, we fabricate four samples corresponding to above connecting schemes and place an acoustic point source on the right side [Fig. 9(a)] of a cavity to excite an acoustic pressure field. We then detect the pressure responses at four typical positions. In the middle panels of Fig. 9, we see the appearance of two resonance peaks in the measured amplitude spectra in the range of 2.3 to 2.6 kHz near the eigenfrequencies (grey arrows). For the in-phase (out-of-phase) mode, positions A and C or B and D have 0 ($\pi$) phase difference. The phase spectra in the lower panels of Figs. 9(a) and 9(c) [Figs. 9(b) and 9(d)] illustrate that the in-phase mode has lower (higher) frequency than the out-of-phase mode, indicating that we implement the negative (positive) coupling $\kappa$.

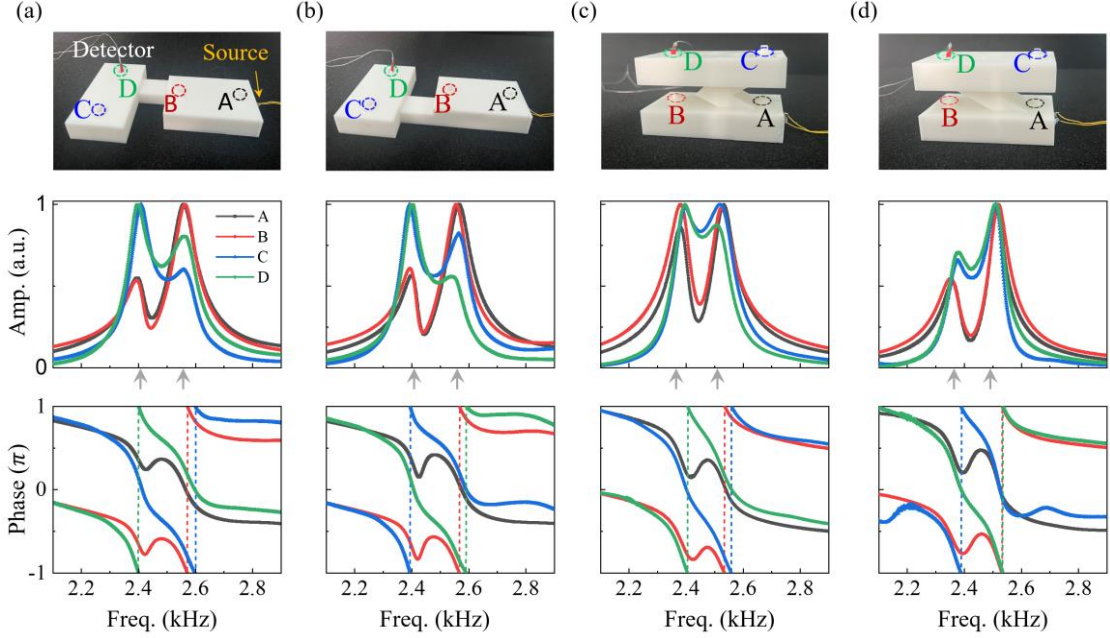

FIG. 9. Demonstration of positive and negative couplings between two acoustic cavities. (a), (b) Negative and positive couplings in the *x-y* plane. (c), (d) Negative and positive couplings along the $z$ direction. Upper panels: pictures of the two coupled acoustic cavities and experimental setups. An acoustic point source is placed on the right side of a cavity to excite an acoustic pressure field, and the pressure responses are detected at four typical positions marked out as A, B, C and D. The two cavities are coupled by a narrow tube with different geometric configurations to realize either positive or negative couplings. Middle panels: measured amplitude responses at four positions A (black), B (red), C (blue), and D (green). The grey arrows point out the numerically calculated resonant frequencies. Lower panels: the corresponding measured phase spectra.

## 2. Effects of loss and measured transmission spectrum

In this section, we will discuss the effect of loss on the acoustic propagation and present the measured transmission spectrum.

In an acoustic experiment, the loss during propagation of acoustic waves is inevitable. Hence, in the simulation, we introduce the complex acoustic velocity given by $v = 343 * (1 + \alpha i)$ m/s with $\alpha = 0.007$ to describe the system losses. In the main text, we show that the measured results agree well with the simulated ones for the pressure field distributions. We now plot the loss of the acoustic waves as a function of the layer index at the frequency of 2.5 kHz in Fig. 10(a). We see that the experimental results are consistent with the simulated ones. It illustrates that the acoustic wave intensity undergoes about 15dB attenuation along the z direction after propagating through 16 layers. The average attenuation rate is thus less than 1dB, which is close to the measurements of hinge states in other papers [34].

In addition, we present the measured transmission spectrum by placing a source at the bottom of the sample and detecting the signal at the top surface. To measure the bulk transmission spectrum, the source and detector are placed at the center of bottom and top surfaces, respectively. The spectrum exhibits an obvious bandgap 2.47-2.54kHz (grey region) as shown by the black line in Fig. 10(b). To measure the hinge transmission spectrum, we place the source and detector at the corners of bottom and top surfaces, respectively. The measured transmission exhibits a significant increase in the frequency range of 2.36-2.65 kHz, providing additional evidence for the existence of hinge states.

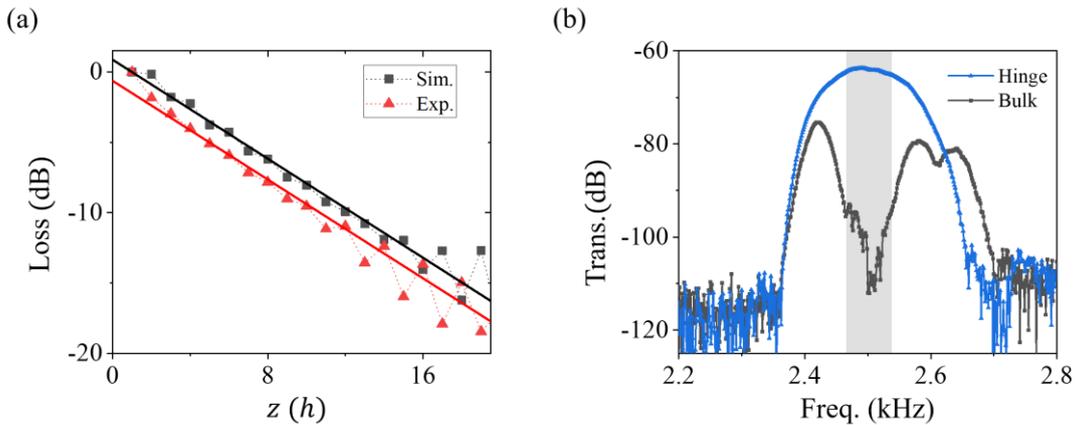

FIG. 10. (a) Loss of the acoustic waves along the $z$ direction. Red triangles and black

squares denote the experimental and simulated results (which are normalized by their maximum values), respectively. Their fitting curves (solid lines) have almost the same slopes, indicating consistent attenuation behavior. (b) The measured transmission spectrum of hinge (blue) and bulk (black) states by putting the source at the corner and center of bottom surface and detector at the corner and center of the top surface, respectively. The grey region highlights the bulk band gap.

### 3. Measured acoustic pressure field distribution in the $x$-$y$ plane

In Fig. 4(c) of the main text, we have shown the measured acoustic pressure field distribution on the surface. In this section, we present the measured field distribution in the $x$-$y$ plane at the 11th layer in Fig. 11(a). We see the localization of the field distribution near a corner, which agrees well with the simulated eigen field distribution shown in Fig. 11(b). The results provide additional evidence for the existence of hinge states.

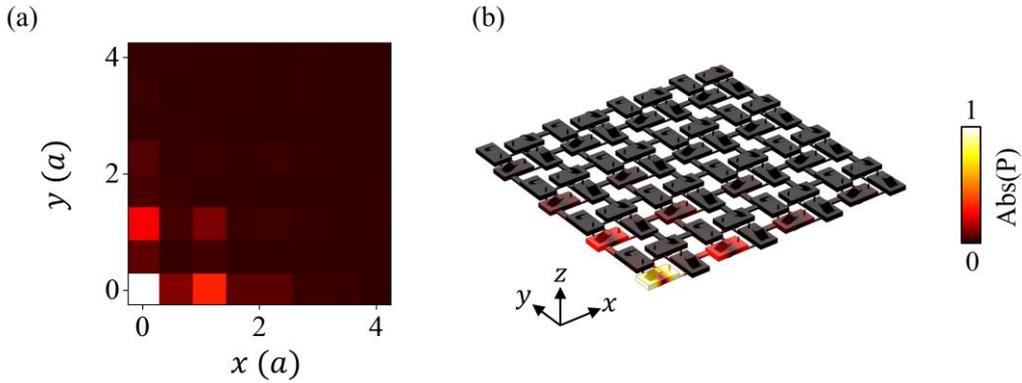

FIG. 11. (a) Measured and (b) simulated pressure field distributions of the hinge states in the $x$-$y$ plane in Fig. 4(c) at the frequency of 2.5 kHz. The fields are normalized to their maximum.

### 4. Measured acoustic pressure field distributions at other frequencies

In Fig. 4(c) of the main text, we have shown the measured acoustic pressure field distribution when an acoustic source with the frequency of 2.50 kHz is placed at the position $S_2$ [see Fig. 3(a)]. In this section, we further present the measured acoustic pressure field distributions at the frequencies of 2.34 and 2.68 kHz by placing the source at the same position in Fig. 12. The figure illustrates that the acoustic wave cannot propagate along the hinge due to the exciting frequency beyond the frequency range (2.36-2.65 kHz) of the typical hinge states. The results are in contrast to the result in Fig. 4(c) of the main text at the frequency of 2.50 kHz, where the propagation of the wave is observed due to the existence of hinge states at this frequency.

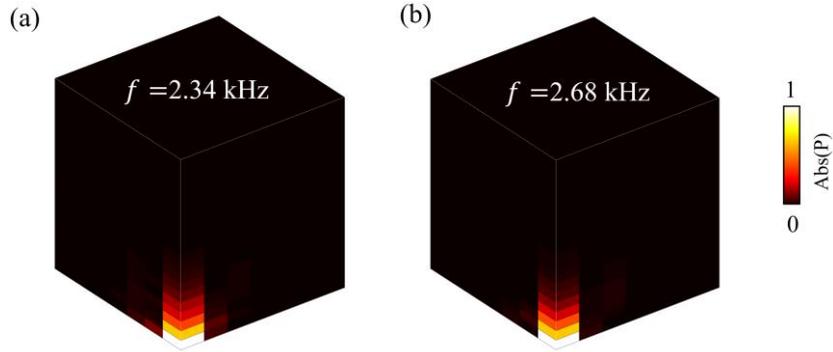

FIG. 12. Measured acoustic pressure field distributions at the frequency of (a) 2.34 kHz and (b) 2.68 kHz.

## APPENDIX F: EFFECTS OF WEAK DISORDER

In this section, we will study the effect of weak disorder by introducing onsite disorder which preserves momentum-space glide reflection symmetries.

The disorder term is expressed as

$$H_\mathrm{d} = W \sum_{r,\alpha} h_r |r,\alpha\rangle\langle r,\alpha|, \quad (S58)$$

where $h_r$ is a random variable which is uniformly distributed in the range of $[-1,1]$, and $W$ is the disorder strength. To ensure that the topology is well defined, we require that $\{h_r\}$ respect the two reflection symmetries. Figure 13(a) displays the energy spectrum of one random sample by considering the disorder without breaking the translational symmetry along $z$. Clearly, gapless hinge modes persist, indicating the stability of the topologically nontrivial phase against weak disorder. Figure 13(b) further plots the configuration averaged $\chi_q$ as a function of $W$, showing that the phase is stable against weak disorder. In addition, we consider the disorder in 3Ds that break the translational symmetry and find that gapless states at zero energy are mainly localized at the hinges as revealed by the local density of state in Fig. 13(c), further indicating the stability of our phase against weak disorder.

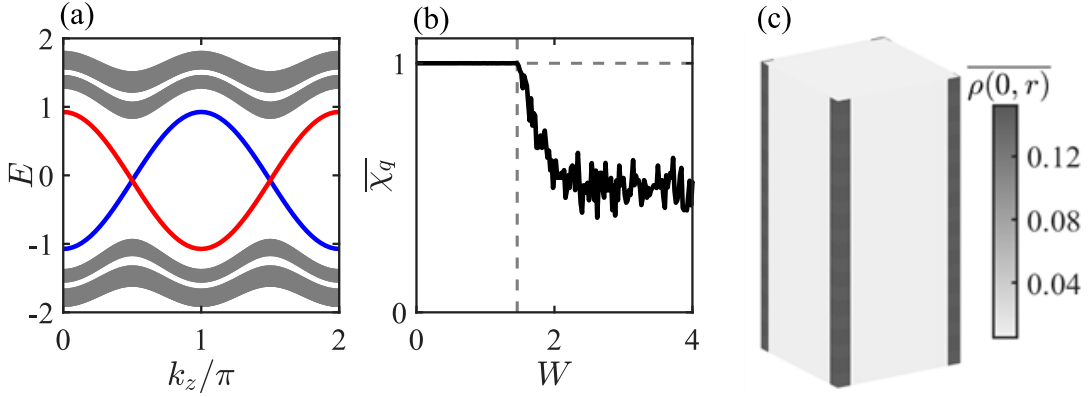

FIG. 13. (a) The energy spectrum versus $k_z$ for the Hamiltonian in Eq. (3) in the main text including the onsite disorder term $H_\mathrm{d}$ under open boundary conditions in the $x$-$y$ plane. Here we only consider one random configuration with $W = 0.05$. The blue and red lines denote the hinge modes localized at diagonal and off-diagonal hinges, respectively. (b) The configuration averaged $\overline{\chi_q}$ as a function of $W$. $\overline{\chi_q}$ suddenly drops from 1 at $W \approx 1.46$ (dotted line), indicating the occurrence of a phase transition. (c) The configuration averaged local density of state $\overline{\rho(E,\boldsymbol{r})}$ at zero energy for disordered systems with $W = 0.05$ under open boundary conditions in the $x$-$y$ plane. The distribution of disorder respects two reflection symmetries. Here, $t_x = t_y = 0.1$ and the size in the $x$-$y$ plane is $10 \times 10$. The number of random configurations in (b) and (c) is 100. We also set $a_x = a_y = a_z = 1$.

## APPENDIX G: ANOMALOUS HIGHER-ORDER KLEIN BOTTLE TOPOLOGICAL INSULATORS

In this section, we will introduce a new 3D tight-binding model (respecting momentum-space nonsymmorphic symmetries) that exhibits an anomalous higher-order Klein bottle topological insulator that harbors gapless hinge modes only in the entanglement spectrum. We write down its Bloch Hamiltonian in momentum space as

$$\mathcal{H}_3(\boldsymbol{k}) = -t_x \tau_3 \sigma_1 + t_y \tau_1 \sigma_1 + t'_x \cos k_x\, \tau_0 \sigma_1 - t'_x \sin k_x\, \tau_3 \sigma_2 \qquad (S59)$$

$$+t'_y \cos k_y \, \tau_2\sigma_2 + t'_y \sin k_y \, \tau_1\sigma_2 + 2t_z \cos k_z \, \tau_0\sigma_3,$$

where we have set $a_x = a_y = a_z = 1$. The Hamiltonian respects two glide reflection symmetries $m_x = \tau_0\sigma_1$ and $m_y = \tau_2\sigma_2$, i.e., $m_x \mathcal{H}_3(\boldsymbol{k}) m_x^{-1} = \mathcal{H}_3(-k_x, k_y + \pi, k_z + \pi)$ and $m_y \mathcal{H}_3(\boldsymbol{k}) m_y^{-1} = \mathcal{H}_3(k_x + \pi, -k_y, k_z + \pi)$. Here we set $t'_x = t'_y = 1$ and $t_z = 0.5$.

Although the glide momenta are different from those of the original glide reflection symmetries, we can also prove that the relation of the quadrupole moment in Eq. (S20) remains valid. For the new $\mathbb{Z}_2$ gauge field, we need to introduce new gauge transformations, and the action of reflection operators on the creation and annihilation operators expressed as Eq. (S23)-(S26) is changed to

$$\widehat{\mathcal{M}}_x \hat{c}^\dagger_{r,\alpha,k_z} \widehat{\mathcal{M}}_x^{-1} = (-1)^y \hat{c}^\dagger_{D_{\widehat{\mathcal{M}}_x}(r,\alpha), k_z+\pi}, \tag{S60}$$

$$\widehat{\mathcal{M}}_x \hat{c}_{r,\alpha,k_z} \widehat{\mathcal{M}}_x^{-1} = (-1)^y \hat{c}_{D_{\widehat{\mathcal{M}}_x}(r,\alpha), k_z+\pi}, \tag{S61}$$

$$\widehat{\mathcal{M}}_y \hat{c}^\dagger_{r,\alpha,k_z} \widehat{\mathcal{M}}_y^{-1} = (-1)^x (-1)^{\theta_\alpha} \hat{c}^\dagger_{D_{\widehat{\mathcal{M}}_y}(r,\alpha), k_z+\pi}, \tag{S62}$$

$$\widehat{\mathcal{M}}_y \hat{c}_{r,\alpha,k_z} \widehat{\mathcal{M}}_y^{-1} = (-1)^x (-1)^{\theta_\alpha} \hat{c}_{D_{\widehat{\mathcal{M}}_y}(r,\alpha), k_z+\pi}, \tag{S63}$$

where $\theta_\alpha$ takes the value of 0 or 1. Clearly, the new gauge transformations have no effects on $\hat{c}^\dagger_{r,\alpha,k_z} \hat{c}_{r,\alpha,k_z}$, so Eq. (S29) and Eq. (S31) remain unchanged. As a result, Eq. (S20) remains valid, and we can utilize $\chi_q$ to characterize the higher-order topology in this new model.

Figure 14(a) displays the phase diagram as a function of $t_x$ and $t_y$ based on $\chi_q$. We find that in the circle region $t^2 = t_x^2 + t_y^2 < 2$, $\chi_q = 1$, indicating that the system is in a topologically nontrivial phase. Remarkably, gapless hinge modes in the energy spectrum only arise in the light red region with $|t_x| < 1$ and $|t_y| < 1$. For the particular dark grey regions, no gapless hinge modes exist in the energy spectrum [see Fig. 14(b) with $t_x = 1.2$ and $t_y = 0.1$]. Based on Ref. [48], we utilize entanglement spectrum to characterize the higher-order topology. The entanglement spectrum is determined by eigenvalues of the correlation matrix in a quarter subsystem $A$ in the $x$-$y$ plane [see Fig. 14(b)] defined as [54]

$$[C_A(k_z)]_{r_i\alpha, r_j\beta} = \langle \psi_G | \hat{c}^\dagger_{r_i,\alpha,k_z} \hat{c}_{r_j,\beta,k_z} | \psi_G \rangle, \qquad (S64)$$

where $|\psi_G\rangle$ denotes the many-body ground state of the Hamiltonian in Eq. (S59) at half filling. Fig. 14(d) displays the entanglement spectrum as a function of $k_z$ with the same parameters in Fig. 14(c), and we can find that nontrivial gapless hinge modes appear in the entanglement spectrum. In fact, one can observe gapless hinge modes in the entanglement spectrum for the whole circle region.

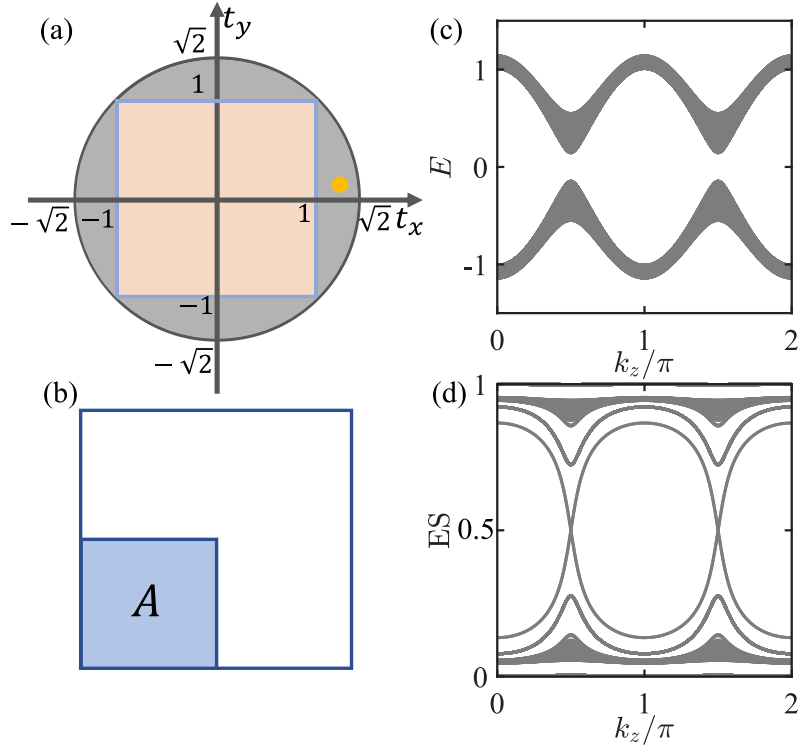

FIG. 14. (a) The phase diagram as a function of $t_x$ and $t_y$ for the Hamiltonian in Eq. (S59). The light red region represents the topologically nontrivial phase with $\chi_q = 1$ hosting gapless hinge modes in the energy spectrum. The dark grey regions depict the topologically nontrivial phase with $\chi_q = 1$ supporting gapless hinge modes only in the entanglement spectrum. (b) Schematics for calculating the entanglement spectrum in the quarter subsystem $A$ in the $x$-$y$ plane. (c) The energy spectrum and (d) the entanglement spectrum with respect to $k_z$ for the Hamiltonian with $t_x = 1.2$ and $t_y = 0.1$ [marked out as a solid yellow circle in (a)].